\documentclass[runningheads]{llncs}

\usepackage{wrapfig}

\makeatletter
\def\orcidID#1{\smash{\href{https://orcid.org/#1}{\protect\raisebox{-1.25pt}{\protect\includegraphics{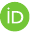}}}}}
\makeatother

\usepackage[colorlinks]{hyperref}
\let\oldparagraph\paragraph
\newcommand{\denseparagraph}[1]{\vspace*{-2.7mm}\oldparagraph{#1}}
\let\paragraph\denseparagraph

\usepackage{xcolor}

\definecolor{koehlma-blue}{RGB}{61,107,181}
\definecolor{koehlma-turquoise}{RGB}{70,180,177}

\hypersetup{
  linkcolor=blue,
  citecolor=blue,
  urlcolor=blue,
  filecolor=blue
}

\usepackage{xspace}
\usepackage{textcomp}
\usepackage{cleveref}
\usepackage{wrapfig}
\usepackage[per-mode=symbol]{siunitx}
\usepackage{listings}
\usepackage{booktabs}
\usepackage{caption}
\usepackage{subcaption}
\usepackage{makecell}
\usepackage{marvosym}
\def\euro{\EUR}

\usepackage[firstpage]{draftwatermark}
\SetWatermarkText{\hspace*{4.5in}\raisebox{6.3in}{\includegraphics[scale=0.1]{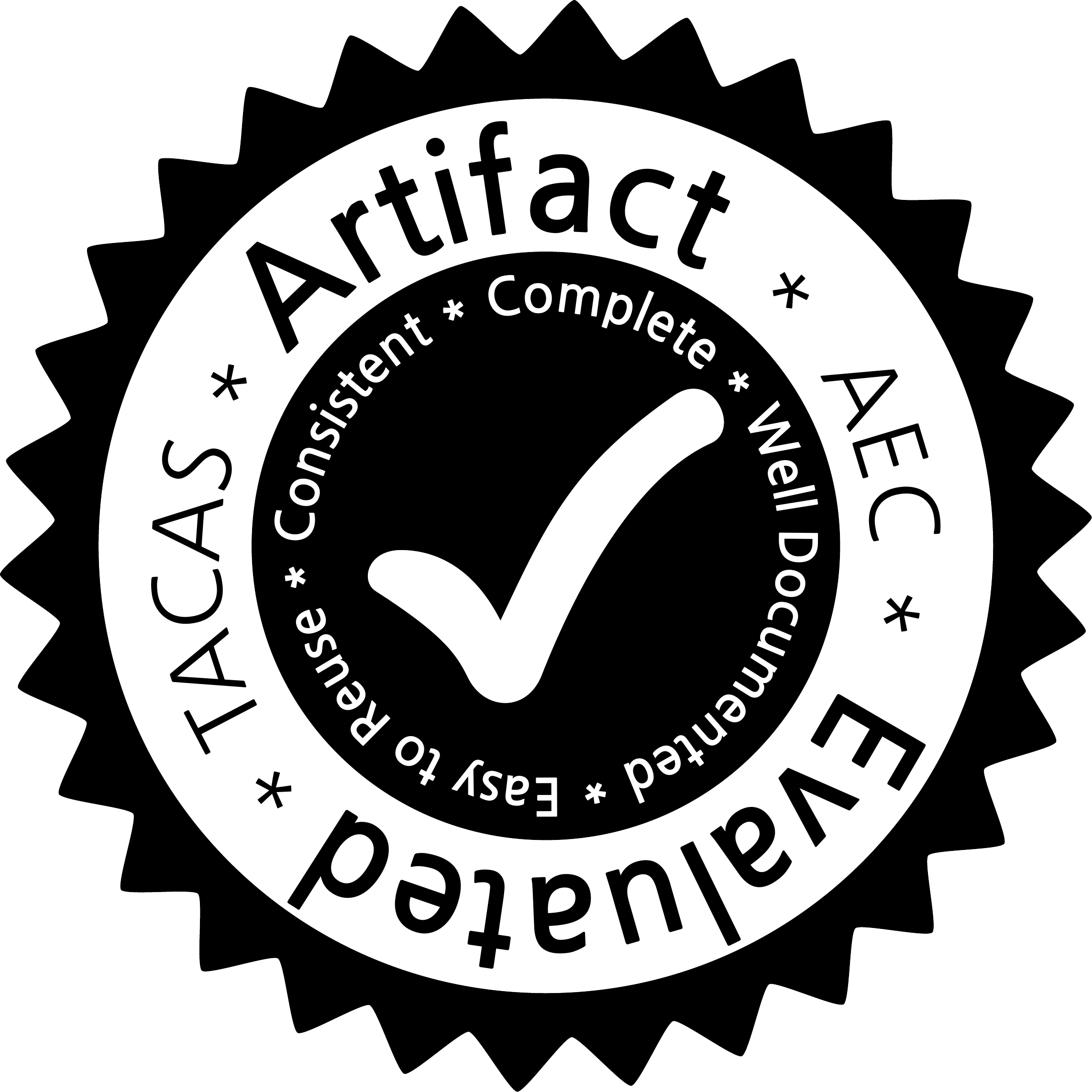}}}
\SetWatermarkAngle{0}

\newcommand{\NOx}{\ensuremath{\mathrm{NO}_x}\xspace}
\newcommand{\COx}{\ensuremath{\mathrm{CO}_2}\xspace}
\newcommand{\androidlola}{\APP}
\newcommand*{\eg}{e.g.\@\xspace}
\newcommand{\APP}{\textsc{LolaDrives}\xspace} 

\newcommand{\rtlola}{\textsc{rtlola}\xspace}
\newcommand{\obd}{\textsc{obd}\xspace}
\newcommand{\ui}{\textsc{ui}\xspace}
\newcommand{\rde}{\textsc{rde}\xspace}

\newcommand{\pems}{\textsc{pems}\xspace}
\newcommand{\pcdp}{\textsc{cdp}\xspace}

\definecolor{bluekeywords}{rgb}{0.13, 0.13, 1}
\definecolor{greentypes}{rgb}{0, 0.5, 0}
\definecolor{redstrings}{RGB}{171, 114, 2}
\definecolor{graynumbers}{rgb}{0.5, 0.5, 0.5}
\definecolor{goldcomments}{rgb}{0.6, 0.4, 0.08}
\definecolor{monitorblue}{RGB}{18, 163, 38}

\lstdefinelanguage{Lola}{
  keywords=[0]{input, output, trigger, import},
  keywordstyle=[0]\bfseries\color{bluekeywords},
  keywords=[1]{if, then, else, aggregate, defaults, offset, filter, hold},
  keywords=[2]{Int8, Int16, Int32, Int64, UInt8, UInt16, UInt32, UInt64, Bool, Float16, Float32, Float64, @1Hz, @2Hz, @5Hz, @10Hz, @100mHz, @1kHz},
  keywordstyle=[2]\color{greentypes},
    sensitive=false,
    comment=[l]{//},
    morecomment=[s]{/*}{*/},
    morestring=[b]',
    morestring=[b]"
}

\begin{document}
\title{RTLola on Board: Testing Real Driving Emissions on your Phone\thanks{This work is partly supported by DFG grant 389792660 as part of \href{https://perspicuous-computing.science}{TRR~248 -- CPEC}, by the European Research Council~(ERC) grants 683300 (OSARES), 695614 (POWVER), and 966770~(LEOpowver), and by the Key-Area Research and Development Program Grant 2018B010107004 of Guangdong Province.}}

%
%
\author{Sebastian Biewer\inst{1}\textsuperscript{(\Letter)}  \orcidID{0000-0002-6897-2506}
  \and
  Bernd Finkbeiner\inst{2} \orcidID{0000-0002-4280-8441}
  \and \\
  Holger Hermanns\inst{1,3} \orcidID{0000-0002-2766-9615}
  \and
  Maximilian A. K\"ohl\inst{1} \orcidID{0000-0003-2551-2814}
  \and \\
  Yannik Schnitzer\inst{1} \orcidID{0000-0001-7406-3440}
  \and
  Maximilian Schwenger\inst{2} \orcidID{0000-0002-2091-7575}
}
\authorrunning{S. Biewer et al.}
%
\institute{Saarland University, Saarland Informatics Campus, Saarbr\"ucken, Germany \email{biewer@depend.uni-saarland.de} \and
CISPA Helmholtz Center for Information Security, Saarbr\"ucken, Germany \and
Institute of Intelligent Software, Guangzhou, China}
\maketitle              

\begin{abstract}
  This paper is about shipping runtime verification to the masses. It
  presents the crucial technology enabling everyday car owners to
  monitor the behaviour of their cars
  in-the-wild. Concretely, we present an Android app that deploys
  \rtlola runtime monitors for the purpose of diagnosing
  automotive exhaust emissions. For this, it harvests the availability
  of cheap bluetooth adapters to the On-Board-Diagnostics~(\obd) ports, which are ubiquitous in cars nowadays. 
  We detail its use in the context of Real Driving Emissions (\rde) tests and report on sample runs that helped identify violations of the regulatory framework currently valid in the~European Union.  
\end{abstract}
%
  

%
%
%
\section{Introduction}
\label{sec:introduction}
\begin{wrapfigure}[0]{r}{0.25\linewidth}
\vspace{-22mm}
\end{wrapfigure}
In the last decade, far more than 600 million cars have
entered the streets worldwide~\cite{car-sales}. 
With very few exceptions, each of these
cars is equipped with a standardized On-Board-Diagnostics (\obd~\cite{OBD2})
interface. Five years ago it surfaced that many of the cars out there
do not adhere to the regulatory framework with which they are supposed to comply. 
For example, a number of undeniable proofs of tampered emission
cleaning systems in passenger cars~\cite{DBLP:conf/sp/ContagLPDLHS17,stadler-prison,VW-fine} are known by now. When
this scandal first surfaced, the regulations imposed by the
authorities were related to isolated tests carried out under lab-like
conditions on chassis dynamometers~\cite{nedc,DBLP:conf/qest/BiewerDH19}. Since then, there has been a
growing understanding that emission and fuel or battery consumption measurements should best take place in a realistic context. Hence, the first test framework for testing on public
roads, the \emph{Real Driving Emissions} (\rde) test has been
developed~\cite{TUTUIANU201561,LEX:32017R1151} and is being rolled out for car model 
approval  in Europe and other entities of jurisdiction.

The \rde regulation specifies the conditions under which a car trip
qualifies as a valid \rde test. These conditions refer to the
trajectory driven, duration, altitudes, speeds, and on the dynamics
of the driving profile~\cite{LEX:32017R1151}. By combining the information available at the
\obd port and the position of the car, it is possible to cast \rde
testing into a runtime monitoring~\cite{watanabe,rtutjournal,lee} problem. Indeed we have shown in
earlier work~\cite{hermanns:2018:verification} how to formalize the \rde 
regulations in \rtlola~\cite{rtlolacavtoolpaper,rtlolaarxiv}, a real-time extension of the stream-based specification language Lola~\cite{lola}. Lola combines the ease-of-use of rule-based specification languages with the expressive power of heavy-weight scripting languages or temporal logics. The eponymous framework generates runtime monitors for such specifications, which were successfully deployed, for instance, on unmanned aircraft~\cite{Torens2017,rtlolacavindustrial}.

An official \rde test requires a calibrated \emph{portable emissions measurement system} (\pems) to be connected to the car's
exhaust pipe while driving the test, so as to correctly quantify the amount of
exhaust emissions induced. The purchasing costs
of a \pems are in the order of \euro{250,000} which is close to unaffordable even in a research context. 
However, many car models expose a variety of diagnosis data through \obd and an \obd-to-Bluetooth adapter can be purchased for around \euro{10}. The data exposed depends on the type of engine, emission cleaning system, and other components in use. 
There are several minimal combinations of \obd data giving good approximations of emitted gases. In particular, various car models expose the sensor readings of their after-treatment \NOx sensor deployed at the rear of the exhaust pipe. 

\begin{wrapfigure}[14]{r}{0.251\linewidth}
\vspace{-16mm}
\centerline{\includegraphics[width=\linewidth]{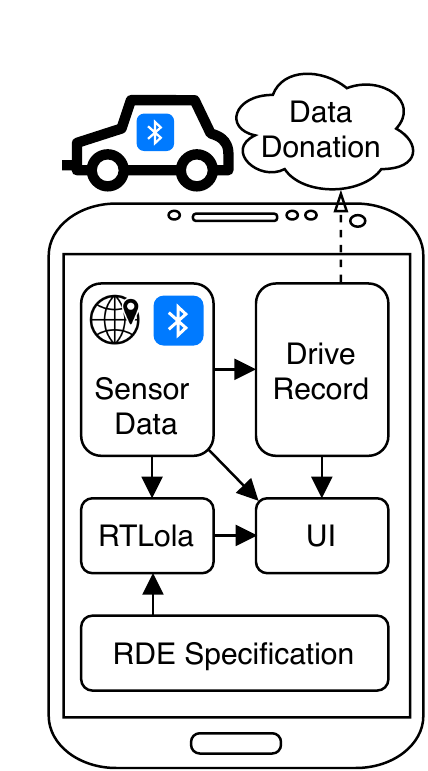}}
\caption{\APP}
\label{fig:overview}
\end{wrapfigure}
\paragraph{Contribution.}
This paper presents \APP, an Android app enabling car owners to carry out real driving emission tests with little investment. Prerequisites are~(i)~an Android phone, (ii)~an \obd-to-Bluetooth adapter, and~(iii)~a car model that does indeed expose the needed values via \obd. If the latter is not the case, the app can still serve the user as a convenient personal monitoring and logging device for the many quantities exposed while driving. 

A structural overview of \APP is depicted in \Cref{fig:overview}. At the
core of the app is an Android version of the \rtlola engine~\cite{rtlolacavtoolpaper}. The engine is strictly separated from the data acquisition and the \rtlola \rde specification. This separation will make it possible to reuse the approach in other runtime monitoring contexts, be it of espresso machines via \textsc{usb}, or drones via Wi-Fi. In both cases, it would especially be the specification in \rtlola that needs to change, not the engine. Car sensor data is acquired via Bluetooth from the \obd device, and combined with location data provided by Android's \textsc{gps} service. 
The data streams are recorded for later diagnosis.
Anticipating future application scenarios involving crowd sourcing car data, we advertise the app as part of a \emph{car data platform} (\pcdp), which includes an upload facility for donating drive records.
While driving, the app's user interface (\ui) displays diagnostic information to the user, both regarding the correct execution of an \rde test drive and the car's emission data. We will detail the separate components of the app next.   

Notably, the lack of any calibration and the unknown precision of the
data exposed by the car manufacturer via \obd make it impossible to
consider the \rde test results reported by \APP as anything more than
indicators of the car's \rde behaviour in a legal sense.

\section{RDE Monitoring on Android}
\label{sec:lola}
The primary feature of \androidlola is to monitor the progress of an \rde test drive.
For this, it uses the \rtlola monitoring framework.
This bridges the gap between formally sound concepts and every-day use cases.
While \rtlola does target a broad audience, that audience is still intended to be expert users rather than the general public.
It requires users to execute three tasks: provide a formal specification of the intended behaviour, supply input data, and interpret the monitor's output.
\androidlola reduces these tasks to minimal action points for end-users.

\paragraph{Specification.}
No end-user input is required with respect to the \rtlola specification. The definition of what is a valid \rde test is fixed~\cite{hermanns:2018:verification} and strictly follows the constraints imposed by the regulation issued by the European Commission~\cite{LEX:32017R1151}. 
These constraints concern the driving behaviour and layout of the route.
Some of them apply universally, \eg, the ambient temperature must range between \SI{273}{\kelvin} and \SI{303}{\kelvin}.
For others, the \rde regulation differentiates three environments:  urban, rural, and motorway with different environments imposing different restrictions on the car, such as an average velocity between 15 and \SI{40}{\kilo\meter\per\hour} in an urban environment.
A \emph{segment} refers to all parts of the test drive in which the car operates in a certain environment.
While segments may be interrupted, each one needs to occupy a specific share of the total distance travelled.

\paragraph{Input Data Provision.}
\androidlola uses sensor readings provided over the \obd interface as input data.
The user only has to plug the \obd-to-Bluetooth adapter in the respective port at (or close to) the dashboard of her car and pair it with her phone.
The car then automatically transmits data to the phone while driving.

\paragraph{Interpretation of Output.}
While driving, \androidlola assists the user in the critical task of satisfying all the constraints that make up a valid \rde. It provides feedback on the driving behaviour indicating which requirements on the test are satisfied to what extent, and which still need attention. 
Furthermore, it evaluates the measured exhaust data and informs the user of whether or not the car violates emission regulations.
Both of these tasks require an online analysis of driving data.
For this analysis, \androidlola uses the \rtlola monitoring framework.


\paragraph{Foundational Underpinning.}
\rtlola~\cite{rtlolaarxiv,rtlolacavtoolpaper} is a stream-based runtime verification framework.
The \rtlola monitor 
analyses sequences of input data to assess whether or not the system complies with the specification.
The specification language has a formal semantics which enables devising provably correct monitoring algorithms~\cite{maxmaster}. 

An \rtlola specification consists of input stream declarations where each input stream corresponds to a source of input data such as the \NOx sensor of the car.
Output stream declarations then spell out how to filter and refine the input data.
For this, \rtlola provides primitives for complex analyses such as sliding window aggregation for common aggregation functions.
Further, the specification contains binary trigger conditions.
The satisfaction of such a condition constitutes a violation of the specification and prompts the monitor to immediately relay a warning to the user.
The following snippet is an extract of an \rtlola specification for \rde test drives~\cite{DBLP:conf/rv/KohlHB18}:
\begin{lstlisting}
input velo_kmph, accel_mpss: Float64
output is_rural := ...      output rural_avg_velo := ...
output rural_dyn : Float64 @1Hz filter: is_rural := velo_kmph * accel_mpss / 3.6
output rural_pctl_dyn : Float64 @1Hz := 
    rural_dyn.aggregate(over: 7200, using: pctl(95)).defaults(to: 0.0)
trigger rural_pctl_dyn > (0.136 * rural_avg_velo + 14.44)
    $\land$ rural_avg_velo <= 74.6
\end{lstlisting}
This specification fragment checks whether the car complies with the \rde regulations regarding the driving dynamics in the rural segment\footnote{See Annex IIIA, Appendix 7a, 3.1.3 in the \textsc{eu} regulations~\cite{LEX:32017R1151}.}.
The first line declares two input streams representing the velocity in \SI[per-mode=reciprocal]{}{\kilo\meter\per\hour} and acceleration in \SI[per-mode=reciprocal]{}{\meter\per\second\squared} supplied by the car.
The third line computes the dynamics in \SI[per-mode=reciprocal]{}{\meter\squared\per\second\cubed}, by multiplying the velocity and acceleration. 
The regulations then demand that the 95\textsuperscript{th} percentile of the dynamics are no greater than $0.136 \cdot v_\mathit{avg} + 14.44$ where $v_\mathit{avg}$ is the average velocity of the vehicle.  
The computation of the velocity and the dynamics only consider sensor readings obtained while in the rural segment.
The full specifications are publicly available~\cite{powver-rde}. 
Note that while the specification is relatively easy to design and understand for computer scientists and engineers, it exceeds the expertise expectable of laymen users. 
However, it is not necessary for them to be confronted with the full potential of the language because \androidlola comes preconfigured with a set of \rde-specific specifications.

As can be seen, the requirements on the end-user are minimal.
Thus, the setup enables users to conduct \rde test drives and assess the emission-behaviour of their cars without requiring them to understand the underlying technology.

\begin{figure}[t]
	\begin{minipage}{.31\textwidth}
		\begin{subfigure}[b]{\columnwidth}
			\centering 
			\includegraphics[width=\columnwidth]{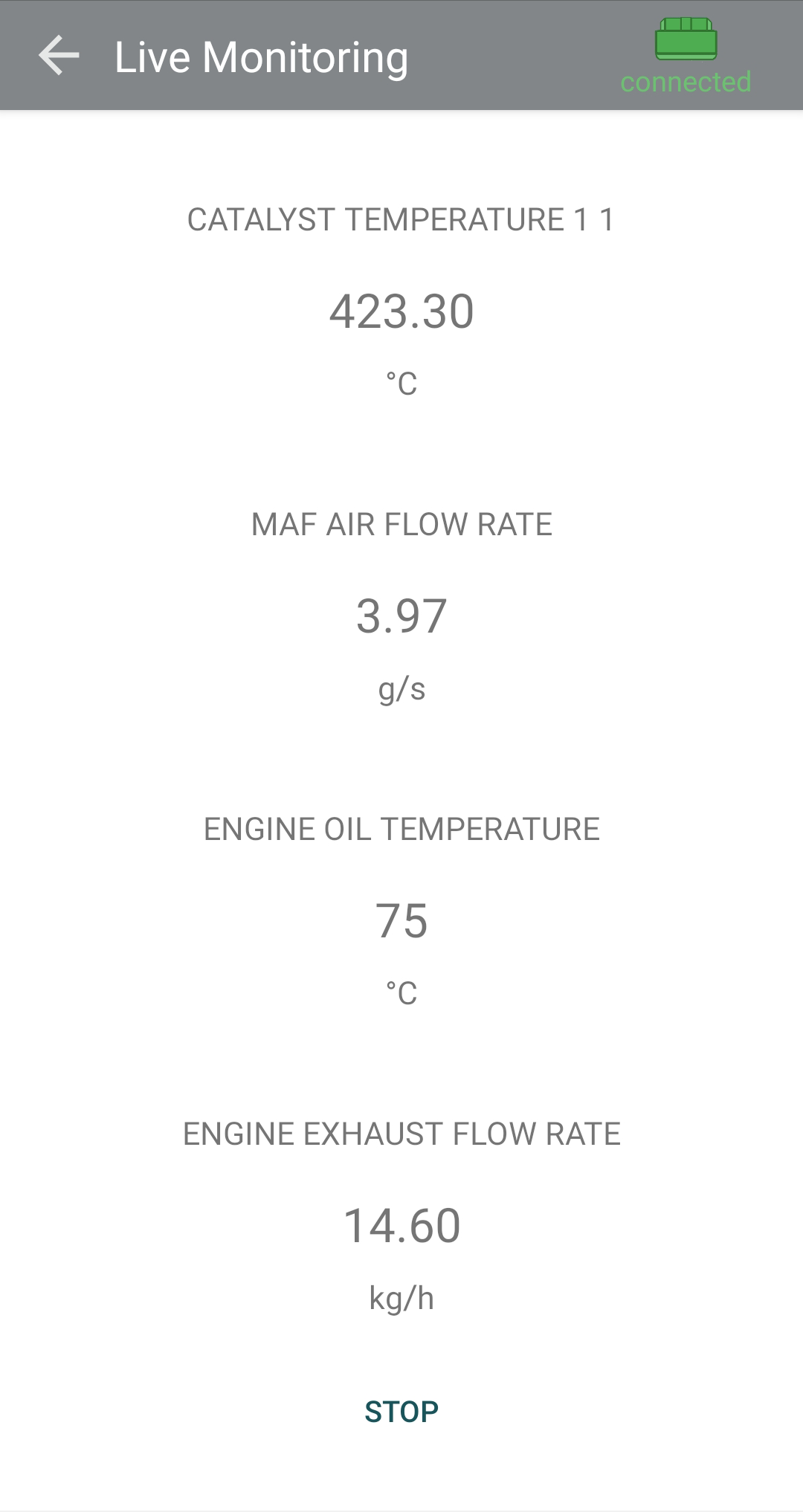}
			\caption{Diagnostics view displays the most recent diagnostics information.}
			\label{fig:diagnosticsview}  
		\end{subfigure}
	\end{minipage}
	\hfill
	\begin{minipage}{.31\textwidth}
		\begin{subfigure}[b]{\columnwidth}
			\centering 
			\includegraphics[width=\columnwidth]{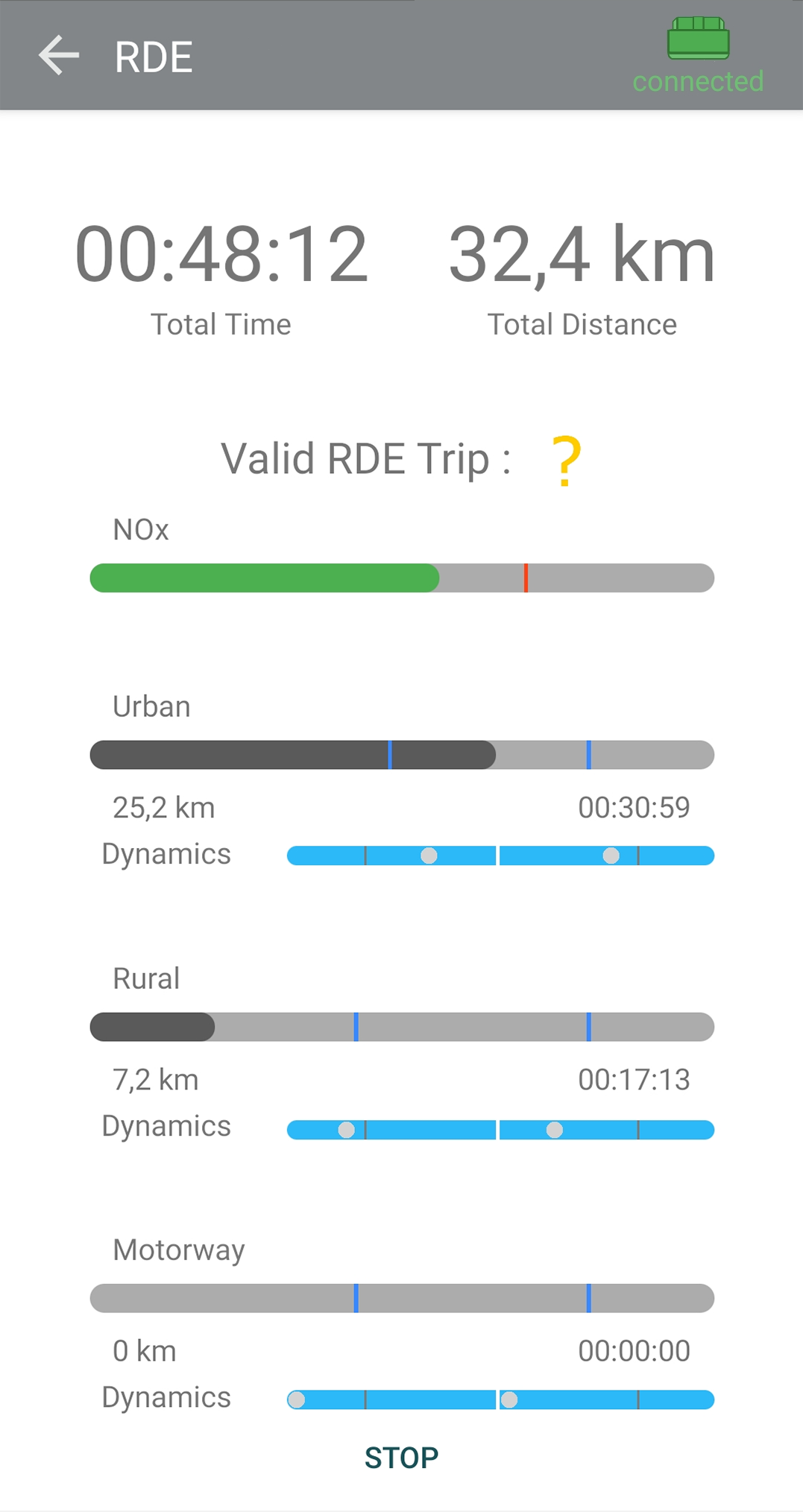}
			\caption{\rde progress view displays current state parameters of the test drive.}
			\label{fig:rdeview}  
		\end{subfigure}
	\end{minipage}
	\hfill
	\begin{minipage}{.31\textwidth}
	    \setlength{\fboxrule}{0pt}%
	    \setlength{\fboxsep}{0pt}%
		\begin{subfigure}[t]{\columnwidth}
			\centering 
			\fcolorbox{white}{white}{\includegraphics[width=\columnwidth]{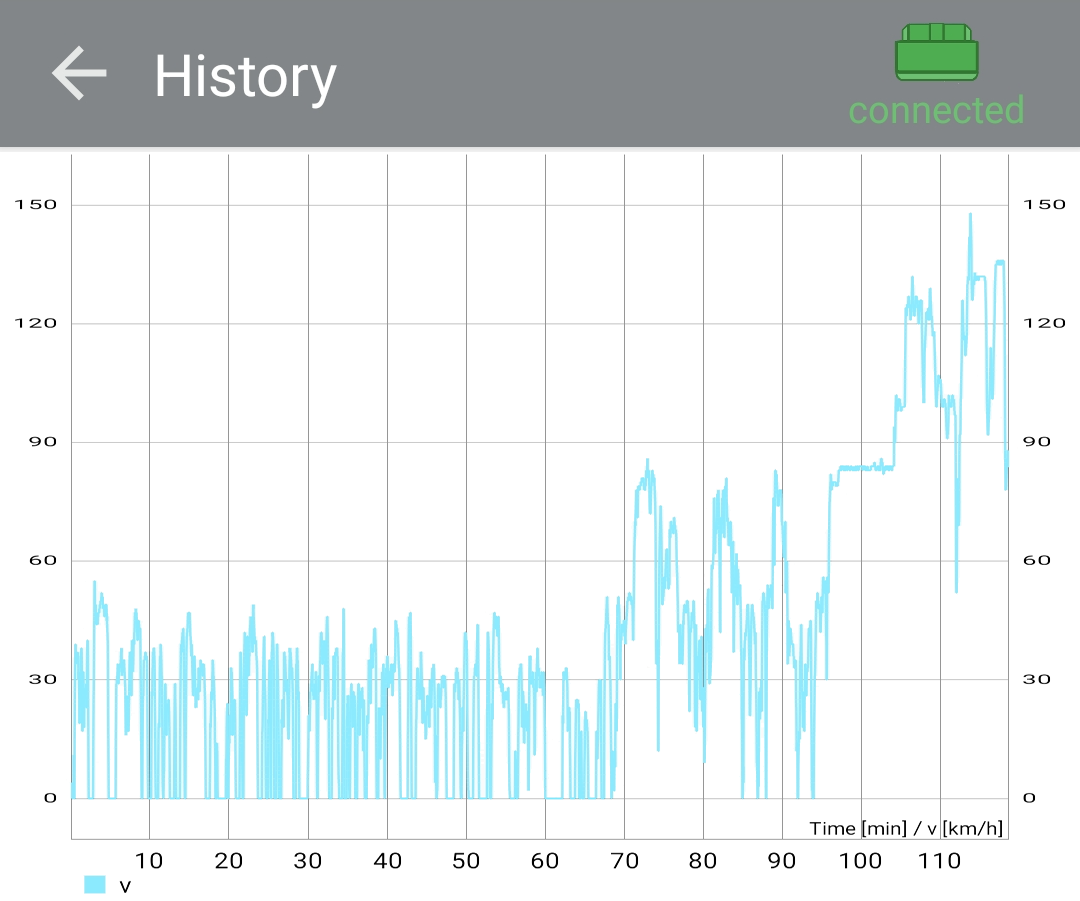}}
			\caption{Replay view displays data plotted against time.\vspace{4.1mm}}
			\label{fig:replayview}  
		\end{subfigure}
		\begin{subfigure}[b]{\columnwidth}
			\centering 
			\fcolorbox{white}{white}{\includegraphics[width=\columnwidth]{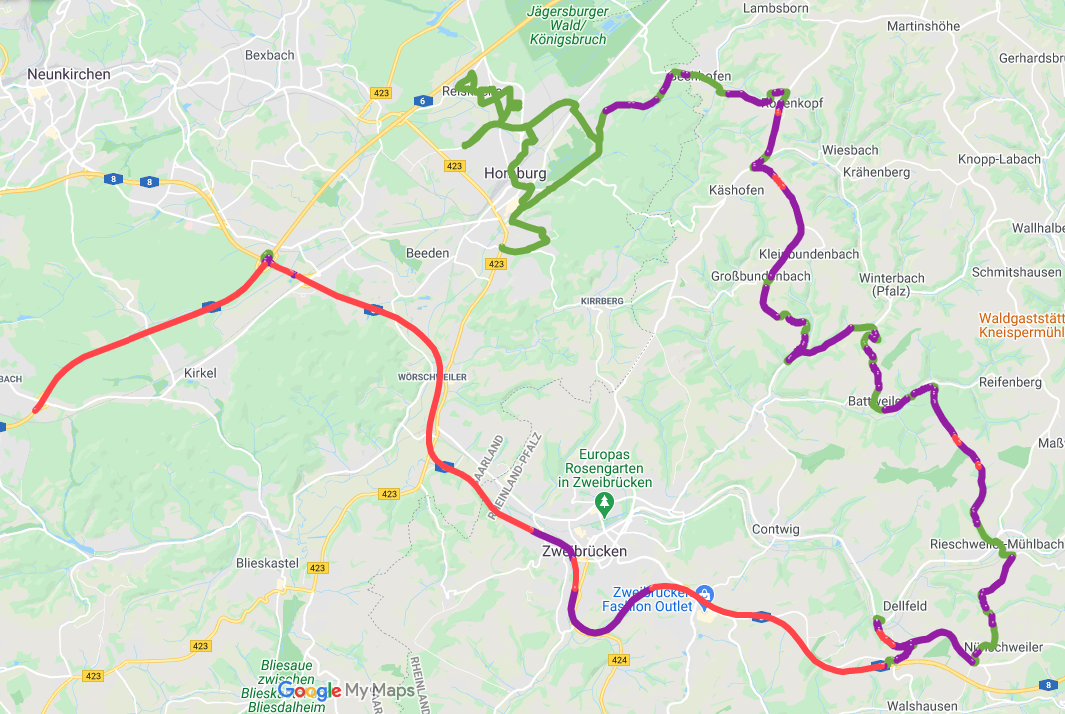}}
			\caption{Map of the second \rde test drive.\vspace{3.9mm}}
			\label{fig:mapview}  
		\end{subfigure}
	\end{minipage}
	\caption{\ui of \APP displaying different views and a map of a test route.}
	\label{fig:screenshots}
	\vspace{-7mm}
\end{figure}

\section{User Experience}
\label{sec:rde}
This section discusses the user perspective on \APP. After a general overview, we report on the use of \androidlola for conducting \rde test drives with a rented vehicle (the precise car model being unknown upfront). 

\paragraph{Overview.}
The preparation of the test requires the user to plug the \obd-adapter
into the \obd-port of the car.
After starting car and app, \androidlola
receives data packets and determines the sensor profile of the car,
assuming phone and adapter are paired via Bluetooth.  
Some sensor profiles provide insufficient data to conduct an \rde test drive.  
In this case, the app is still convenient to use for real-time displaying and logging the available data regardless of \rde regulations, see \Cref{fig:diagnosticsview}. 
If the data suffices, the app selects an appropriate specification and initializes the \rtlola monitor.
\androidlola then starts filtering and visualising the
data output and trigger notifications provided by the monitor.

After successful setup, the \ui switches to an \rde guiding view (\Cref{fig:rdeview}).
From top to bottom, it shows the total time, which must be between 90 and \SI{120}{\minute} to finish the test, and the total distance travelled.
The next line indicates the current state of the conditions for a valid \rde test drive disregarding emission data.
In the screenshot, the drive is still in progress and inconclusive, indicated by the question mark.
Instead, the \ui can also indicate success or failure. The latter verdict can occur far before the time limit is reached, caused by an irrecoverable situation such as transgression of the \SI[per-mode=reciprocal]{160}{\kilo\meter\per\hour} speed limit.
Note that the indicator reports the \emph{current} status if the test drive were to end in this moment. Together with the regulatory constraints, this implies that the current verdict can alternate between success and failure from minute 90 to 120. 
As there is no specific point in time when the test ends, the app continues to compute statistics until the tester manually stops it or the \SI{120}{\minute} mark is reached.
Beneath the status indicator is the green \NOx bar displaying the total \NOx emissions.
The vertical red bar denotes the permitted threshold of \SI[per-mode=reciprocal]{168}{\milli\gram\per\kilo\meter}.

The next three \ui groups represent the progress in each of the distinct segments: urban, rural, and motorway. 
Each group consists of two horizontal bars. The gray progress bar displays the distance covered in the respective segment.
The vertical blue indicators denote lower and upper bounds as per official regulation, for an expected trip length of \SI{83}{\kilo\meter}.
The blue bar below the gray one illustrates two different metrics for the driving dynamics.
Both dots need to remain below/above their thresholds.
A more aggressive acceleration behaviour shifts the dots to the right and a passive driving style to the left. 

\begin{table}[t]
\centering
\begin{tabular}{ l c c c @{\hskip .5cm} c c c }
 & \multicolumn{3}{c}{Drive 1} & \multicolumn{3}{c}{Drive 2} \\
  \cmidrule(r{.4cm}){2-4} \cmidrule(l{-.1cm}){5-7}
 & \makecell[cc]{Distance \\$\left[\SI{}{\kilo\meter}\right]$} & \makecell[cc]{\textit{\NOx} \\ $\left[\SI[per-mode=symbol]{}{\milli\gram\per\kilo\meter}\right]$} & \makecell[cc]{\textit{\COx} \\ $\left[\SI[per-mode=symbol]{}{\gram\per\kilo\meter}\right]$} & \makecell[cc]{Distance \\ $\left[\SI{}{\kilo\meter}\right]$} & \makecell[cc]{\textit{\NOx} \\ $\left[\SI[per-mode=symbol]{}{\milli\gram\per\kilo\meter}\right]$} & \makecell[cc]{\textit{\COx} \\ $\left[\SI[per-mode=symbol]{}{\gram\per\kilo\meter}\right]$} \\
 \midrule
Urban & 35.45 & 137 & 222 & 37.46 & 102 & 251 \\
Rural & 22.33 & 305 & 154 & 27.40 & 90 & 172 \\
Motorway & 26.10 & 241 & 153 & 25.37 & 105 & 175 \\
Total  & 83.88 & 214 & 183 & 90.22 & 99 & 205 \\
    \end{tabular}
    \vspace{.3cm}
    \caption{Aggregation of the emission data based on the \pcdp.}
    \label{tbl:tests}
    \vspace{-10mm}
\end{table}

\paragraph{Test Drive.}
The technical framework and visual feedback of the app were tested in two \rde test drives.
Both tests were conducted with an Audi A6 Avant 45-TDI hybrid diesel, which is approved as Euro 6d-TEMP (DG) with an \NOx threshold of \SI[per-mode=reciprocal]{80}{\milli\gram\per\kilo\meter} under lab conditions and \SI[per-mode=reciprocal]{168}{\milli\gram\per\kilo\meter} for \rde conditions.
Among the diagnosis parameters available within this car are vehicle and engine speed, ambient temperature, engine fuel rate, mass air flow, and two \NOx-sensors---one in front and one behind the emission cleaning system in the exhaust pipe.
With this data, exhaust mass flow and fuel consumption can be computed, from which the total amounts of \NOx and \COx can be derived~\cite{DBLP:conf/rv/KohlHB18}.
In both drives, the driving dynamics were close to the allowed maximum, in the first test below and in the second test above the threshold, so the second test drive did not result in a valid \rde test.
In both cases, the app correctly confirmed the satisfaction and violation of the \rde criteria.
In the first drive, the app reported an average \NOx emission of \SI[per-mode=reciprocal]{214}{\milli\gram\per\kilo\meter}. This constitutes a violation of the regulation.

The app also allows for inspection of the driving data in a plotted form~(\Cref{fig:replayview}).
\Cref{fig:mapview} shows the route of an \rde test drive.  
The first half of the time constituted the urban segment~(green).
The next 30-40\% of the test mainly consisted of the rural segment~(purple) followed by the motorway segment~(red).

\paragraph{Data Harvesting.} For further analysis, data can be uploaded to a cloud storage which is part of the car data platform (\pcdp). This platform provides a uniform way to harvest data by specifying a format for collection, analysis, and exchange of this data. 
\pcdp builds upon a \textsc{json} format (\url{https://json-schema.org/})  
containing timestamped events such as an \obd response, including its raw payload. 
As an example, the data presented in \Cref{tbl:tests} is an aggregation of the \rde test drives mentioned above obtained by post-processing the data.

\section{Conclusion}
\label{sec:conclusion}
\APP pushes runtime verification technology into cars and phones of everyday users. The app is available in \emph{Google Play}~\cite{powver-rde}; a version for iOS is already initiated.
Moreover, the car data platform constitutes a crowd-sourcing initiative for car data with the intention to enable large scale analyses of emission data beyond a single trip and a single car model.

\bibliographystyle{splncs04}
\bibliography{bibliography}

\vfill

{\small\medskip\noindent{\bf Legal Attribution} Android, Google Play and the Google Play logo are trademarks of Google LLC.}

{\small\medskip\noindent{\bf Open Access} This chapter is licensed under the terms of the Creative Commons\break Attribution 4.0 International License (\url{http://creativecommons.org/licenses/by/4.0/}), which permits use, sharing, adaptation, distribution and reproduction in any medium or format, as long as you give appropriate credit to the original author(s) and the source, provide a link to the Creative Commons license and indicate if changes were made.}

{\small \spaceskip .28em plus .1em minus .1em The images or other third party material in this chapter are included in the chapter's Creative Commons license, unless indicated otherwise in a credit line to the material.~If material is not included in the chapter's Creative Commons license and your intended\break use is not permitted by statutory regulation or exceeds the permitted use, you will need to obtain permission directly from the copyright holder.}

\medskip\noindent\includegraphics{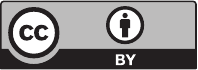}

\end{document}